Enhancement of critical current densities in $(Ba,K)Fe_2As_2$ wires and tapes using HIP technique

Sunseng Pyon[1], Takahiro Suwa[1], Akiyoshi Park[1], Hideki Kajitani[2], Norikiyo Koizumi[2], Yuji Tsuchiya[3,*], Satoshi Awaji[3], Kazuo Watanabe[3], and Tsuyoshi Tamegai[1]

1 Department of Applied Physics, The University of Tokyo, Hongo, Bunkyo-ku, Tokyo 113-8656, Japan
2 Naka Fusion Institute, National Institutes for Quantum and Radiological Science and Technology (QST), 801-1 Mukoyama, Naka-shi, Ibaraki 311-0193, Japan
3. High Field Laboratory for Superconducting Materials, Institute for Materials Research, Tohoku University, Sendai 980-8577, Japan
* Present affiliation: The Department of Energy Engineering and Science, Nagoya University, Nagoya 464-8603, Japan

**Abstract**

$(Ba,K)Fe_2As_2$ superconducting wires and tapes are fabricated by using hot isostatic pressing (HIP) technique, and their superconducting properties are studied. In the HIP round wire, transport critical current density ($J_c$) at 4.2 K has achieved record-high value of 175 kA/cm$^2$ at zero field, and exceeds 20 kA/cm$^2$ even at 100 kOe. Improvement of polycrystalline powder synthesis may play a key role for the enhancement of $J_c$. In the HIP tape, even larger transport $J_c$ of 380 kA/cm$^2$ is realized at zero field. Based on magnetization and magneto-optical measurements, possible further enhancement of $J_c$ is discussed.

**Introduction**

Iron-based superconductors (IBSs) such as $LaFeAsO_{1-x}F_x$ [1] have great potential for the use in high-field applications. They exhibit high critical temperature $T_c$ [2-5], large upper critical field $H_{c2}$ [6-11], and relatively low anisotropy [8,12,13]. It has been demonstrated that the critical current density ($J_c$) in IBS single crystals exceeds $1\times10^6$ A/cm$^2$ even at high magnetic fields. For instance, a high transport $J_c$ over $1\times10^6$ A/cm$^2$ at 5 K under 140 kOe has been reported in a $SmFeAsO_{1-x}F_x$ single crystal [14]. Single crystals of (Ba,K)Fe$_2$As$_2$ also have achieved high $J_c$ over 1x10$^6$ A/cm$^2$ [15-17]. The fact that these values are almost the same level as commercially used NbTi or Nb$_3$Sn wires makes IBSs very promising for high-field applications [18]. Among IBSs, 122-type compounds such as (Ba,K)Fe$_2$As$_2$ and (Sr,K)Fe$_2$As$_2$ [4,19] are studied as promising candidates for high-performance superconducting wires and tapes. Their large $H_{c2}$ and small anisotropies make them attractive compared with MgB$_2$ with $T_c$ ~ 39 K [18,20].

For practical use of superconducting wires, weak links between superconducting grains should be improved. The performance of IBS wires and tapes has been much improved by several methods, such as combination of several times of cold press and hot press [21] or only hot press [16,22-24], only cold press [32,35], addition of metals like Ag, Pb, and Sn [25-31], or by texturing [27,30,31]. Mainly, two pressing methods for the fabrication of IBS wires and tapes have been proven to be effective in enhancing $J_c$, namely, uniaxial pressing for the tape [32-36] and hot isostatic pressing (HIP) for round wires [21,23,39,40]. Although uniaxial-press method has realized the highest $J_c$ value, it cannot be used for the fabrication of round wires. Wires processed by the HIP technique are promising for high-field applications similar to the pressed tapes because their round shapes are much more advantageous and convenient than the textured tape for wide range of applications. It should be noted that the HIP process has been proven to be very effective in enhancing $J_c$ of commercial Bi2223 tapes [37,38]. This also suggests the possibility to apply HIP method for the fabrication of IBS tapes, which has not been attempted yet. Although the HIP method has been applied for the fabrication of 122-type superconducting round wires, the reported values of $J_c$ is not high enough. At present, $J_c$ in both (Ba,K)Fe$_2$As$_2$ and (Sr,K)Fe$_2$As$_2$ HIP wires achieved approximately 100 kA/cm$^2$ at 4.2 K under self-field [21,23,39,40]. However, their $J_c$ at high magnetic fields is reduced to below 20 kA/cm$^2$. These results imply that both polycrystalline powder synthesis and wire fabrication conditions have not been optimized yet.

In this paper, we focus on the HIP technique for the fabrication of (Ba,K)Fe$_2$As$_2$ superconducting wires and tapes. By using a higher quality polycrystalline powder of (Ba,K)Fe$_2$As$_2$, superconducting wires and tapes are fabricated by PIT and HIP method. Obtained HIP wires and tapes have been thoroughly characterized. Both transport and magnetic $J_c$ are measured and compared for both HIP wires and tapes. We have achieved the largest value of $J_c$ in the HIP wire. In addition, a relatively high value of $J_c$ in the HIP tape is also achieved, which suggests that the HIP technique can be an alternative method to realize superconducting tapes with practical $J_c$. Possibility for further enhancement of $J_c$ is discussed based on the analyses of magnetization measurements and magneto-optical imaging.

**Experimental methods**

(Ba,K)Fe$_2$As$_2$ superconducting wires were fabricated by ex-situ powder-in-tube (PIT) method.

Polycrystalline powders of $Ba_{0.6}K_{0.4}Fe_2As_2$ were prepared by the solid-state reaction. Ba pieces, K ingots, Fe powder, and As pieces were used as starting materials. In order to compensate the loss of elements, the starting mixture contained 15% excess K and 5% excess As. They were mixed in a nitrogen atmosphere more than 10 hour using a ball-milling machine and densely packed into a niobium tube. The niobium tube was then put into a stainless steel tube and sealed in nitrogen filled glove box for heat treatment at 900°C for 30 h. The surface of the obtained polycrystalline ingot was polished to remove small impurities. It was then ground into powder using an agate mortar in nitrogen filled glove box. The phase identification of the sample was carried out by means of powder x-ray diffraction (XRD) with Cu-Kα radiation (Smartlab, Rigaku). Ground powders were filled in a silver tube with an outer diameter of 4.5 mm and an inner diameter 3 mm, then cold drawn into a square shape with a diagonal dimension of ~1.2 mm. After cutting it into short pieces, one of the pieces was put into 1/8 inch copper tube and redrawn down to the diagonal dimension of 1.2 mm. Furthermore, a part of the wire was deformed into a tape form with 0.4 mm in thickness. After that, both ends of the wires were sealed by using an arc furnace. The sealed wires were sintered using the HIP technique. Wires and tapes were heated for 4 h at 700°C in argon atmosphere under the pressure of 175 MPa, which were performed by National Institutes for Quantum and Radiological Science and Technology (QST). Bulk magnetization was measured by a superconducting quantum interference device (SQUID) magnetometer (MPMS-5XL, Quantum Design). Current–voltage ($I$–$V$) measurements up to 140 kOe were performed by the four-probe method with solder for contacts. A DC electric current up to 100 A was delivered through the wire at a ramping rate of 50–100 A/min. Measurements were performed in liquid helium to minimize the effect of Joule heating at the current leads. These critical current measurements in high magnetic fields were carried out by using the 15T-SM at High Field Laboratory for Superconducting Materials, IMR, Tohoku University. For Magneto-optical (MO) imaging, the PIT wires were cut and the transverse cross sections were polished with lapping films. An iron-garnet indicator film was placed in direct contact with the sample and the whole assembly was attached to the cold finger of a He-flow cryostat (Microstat-HR, Oxford Instruments). MO images were acquired by using a cooled-CCD camera with 12-bit resolution (ORCA-ER, Hamamatsu).

**Results and discussions**

The quality of polycrystalline powder was examined before the fabrication of wires and tapes by means of powder x-ray diffraction and magnetization measurements. The x-ray diffraction pattern has strong peaks of 122 phase and do not show peaks of impurity phases such as FeAs. In magnetization measurement, a sharp drop of magnetization was detected at 38.2 K, which is the optimal value of $T_c$ in $(Ba,K)Fe_2As_2$. These results guarantee that the quality of the powder sample is sufficiently high, from which we have fabricated wires and tapes. Figure 1 shows the temperature dependence of magnetization for both the HIP wire and tape. $T_c$ in both the HIP wire and tape is approximately 36 K. The small reduction of $T_c$ after the wire fabrication is almost the same compared with the previous reports on the HIP wires with the highest $J_c$ [21,39]. This $T_c$ reduction is caused by the degradation in the drawing process and incomplete recovery by the successive sintering, as discussed in ref [39]. Shielding volume fractions of the HIP wire and tape are roughly 100% and 1300%, respectively, at 10 K. The overestimated volume fraction of the HIP tape over 100% is caused by the demagnetization effect of the tape shape. Relatively high $T_c$, sharp transition width, and large volume fraction indicated that the carrier doping level is uniform.

Transport $J_c$ in these HIP wires and tapes are evaluated from their $E$-$J$ characteristics under various magnetic fields in liquid helium, at $T$ = 4.2 K. We evaluate the transport $J_c$ for the HIP wire by adopting 1 μV/cm criterion. Evaluated transport $J_c$ under magnetic fields is summarized in figure 2. Transport $J_c$ in the $(Ba,K)Fe_2As_2$ HIP wire from recent publications are also plotted for comparison [21,39,40]. Noteworthy achievements are the highest transport $J_c$ value among IBS superconducting HIP wires for both self-field and high magnetic field of 100 kOe. As shown in figure 2(a), the self-field transport $J_c$ reaches 175 kA/cm$^2$. Furthermore, $J_c$ under magnetic field of 100 kOe exceeds 20 kA/cm$^2$ for the first time. The HIP tape also shows high transport $J_c$ as shown in figure 2(b). $J_c$ of the HIP tape is higher than that of the HIP wire. The transport $J_c$ of the HIP tape has reached 254 kA/cm$^2$ in self-field and 23 kA/cm$^2$ in magnetic field of 100 kOe.

Next, we perform MO measurements on the HIP wire and tape. Using MO technique, local $J_c$ distribution in the core can be evaluated as shown below. Figures 3(a) and (b) are optical images of the transverse cross sections of the HIP wire and tape, respectively. Copper, silver, and $(Ba,K)Fe_2As_2$ core regions are clearly identified. No voids are observed in these optical images. The shape of the core of the HIP wire is nearly circular. In the case of the HIP tape, a gourd-shaped core, where the center is thinner than the outer parts, is observed. Figures 3(c) and (d) depict MO images of the transverse cross section of the core of the HIP wire and tape in the remanent magnetization, respectively. These images are obtained after applying 1.5 kOe along the wire axis for 0.2 sec at 5 K, which is subsequently reduced to zero. The bright region corresponds to the trapped flux in the sample. Figures 3(c) shows the uniform and fully trapped magnetic flux distribution in the HIP wire. It demonstrated the presence of uniform bulk current flow in the wire core across many grains, and weak links across grain boundaries are much improved by the HIP process and the intergranular current density is enhanced. In the core of the HIP tape, two bright regions are separated as shown in figure 3(d). The uniform and fully trapped field in the outer region is similar to that in the core of the HIP wire. On the other hand, the central part traps less flux compared with the outer parts. It can be caused by the weaker $J_c$ in the central part or shape effect of thinner central part. First we focused on the outer part, and the central part is discussed later. Figures 3(e) and (f) show the magnetic induction profiles along the red line in figures 3(c) and (d), respectively. Figure 3(e) and (f) show that, as the temperature is increased toward $T_c$, the intergranular critical current in the HIP wire and tape decreases only gradually. Such behavior indicates that a uniform bulk current flows in the core across many grains in the temperature range between 5 K and $T_c$. From the comparison between MO images for the pressed and unpressed samples [23, 28], these observations indicate the fact that weak links between grains are improved by sintering using the HIP technique.

Our HIP wire shows the highest value of $J_c$ among other reported round wire of IBS. Quality of polycrystalline powder may play a key role for the enhancement of $J_c$. To clarify this, characteristics of $(Ba,K)Fe_2As_2$ powders used in this work ("new") and in the previous work ("old") in ref. [39] are compared. Temperature dependence of normalized magnetization of "new" and "old" $(Ba,K)Fe_2As_2$ powders are shown in figure 4(a). Although, the onset $T_c$ defined in the inset of figure 4(a) for "new" powder is a little lower than "old" powder, the temperature where magnetization of "new" powder reaches the 90% values of magnetization at 10 K is higher compared with "old" powder. Sharper drop of magnetization near $T_c$ suggests more homogeneous K distribution. We also performed x-ray diffraction analyses for both powders. Both patterns do not show clear impurity peaks as shown in figure 4(b). However, distinct difference can be seen in the peak width. The inset for figure 4(b) shows the normalized (1 1 6) peaks from the two powders. The peak width for "new" powder is much narrower than that for "old" powder. It is well established that lattice constants in $(Ba,K)Fe_2As_2$ systematically change with K content [41]. So, the smaller peak width also suggests the more homogenous K distribution. Both magnetization and x-ray diffraction measurement suggest the higher quality of "new" powder compared with "old" powder in the previous work [39]. Compared with the method used in the previous report [39], we improved the method of synthesis of polycrystalline powders. When raw materials were mixed by the ball-milling method, not alumina but stainless pot and balls were used. In addition, we used Nb tube instead of alumina tube for synthesis. When ball-milled powder was put in the tube, the powder was pressed into it to promote diffusion of elements during the reaction. After the reaction, outer skin of the reacted material in contact with the Nb tube is removed. These new processes were performed to reduce the impurity phase and to react the raw material sufficiently. It should be noted that only a small amount of impurities such as FeAs can reduce the intergranular $J_c$ considerably [21]. High-quality polycrystalline powder may play a key role to realize the highest $J_c$ in the HIP wire.

For further enhancement of $J_c$, we should also consider the heating temperature for synthesis of powder and sintering wires or tapes. Recently, Hecher *et al*. argued that small grain size is important for achieving high $J_c$ in a round wire [40]. They demonstrate that small-grain polycrystalline powder synthesized at $600^o$C is better than the large-grain polycrystalline powder synthesized at $900^o$C. On the other hand, in our previous report [39], we demonstrate that superconducting property is significantly suppressed by the deformation during the drawing process and it recovers by sintering at higher temperatures. These opposite effects of heating temperature for $J_c$ enhancement suggest the necessity of optimization of heat treatment. To make or keep grain size small and to recover superconducting properties, heating condition for powder synthesis and wire sintering should be investigated. In addition, we believe that microscopic evaluation of grain boundaries using SEM may help for identifying the origin of suppression of intergranular current, such as minor impurities, and removing such origins.

Next, we analyze the HIP tape carefully to discuss whether HIP technique is useful for the fabrication of the tape or not. In this research, although transport $J_c$ of the HIP tape is higher than that of the HIP wire, it is still lower than $J_c$ of pressed tapes [32,33]. As discussed below, however, there may be a room for $J_c$ enhancement by the HIP method. In figure 3, possible weak links in the central part of the core of the HIP tape was suggested. To clarify $J_c$ distribution in the core, we perform MO imaging of the core of the HIP tape. Figure 5 (a) shows an optical image of the transverse cross section of the HIP tape. The shape of the core looks like a gourd with thin central part sandwiched by two thicker outer regions. In order to access the intrinsic $J_c$ properties in the core region, we cut out the thicker parts. An optical image of the polished upper surface is shown in figure 4 5 (b). This process is schematically drawn in figure 5 (c). The final thickness of the tape core is approximately 20 μm. Using this piece, we perform MO measurements. Figures 5 (d) and (e) depict MO images of the polished surface of the core of the HIP tape in the remanent state at 5 K and 25 K, respectively. These images are obtained by the same method as the former measurements shown in figure 3. As shown in figure 5 (d) and (e), left central part shows a relatively small trapped field. This is clearly shown at higher temperature shown in figure 5 (e). Figures 5 (f) shows the magnetic induction profiles along the red line in figures 5 (d). Figure 5 (f) shows that, as the temperature is increased toward $T_c$, dip structures develop in the profiles. The dip structure always appears at the left central part, corresponding the thinnest part of the gourd-shaped core of the HIP tape before polishing in figure 5 (a). It is noteworthy that this dip structure is not caused by microcracks. The inset of figure 5(e) shows MO image of flux penetration in the core of HIP tape at 25 K and 50 Oe after zero-field cooling. Flux penetrations only occur from the edge part, but not into the central part, along tape length. These results suggest that intergranular $J_c$ of the thinner part of HIP core is smaller than the other parts. Such a suppression of $J_c$ in the thinner part of the core may have created by uneven stress during the deformation process. Regardless of its origin, it implies that our fabrication method of the HIP tape have still a room for enhancing global $J_c$ by improving $J_c$ properties in the central part.

Further analyses of $J_c$ in the HIP wire and tape are performed by magnetization measurements. Magnetic $J_c$ of the HIP wire and tape are also evaluated from the irreversible magnetization by using the extended Bean model [39]. Evaluated magnetic $J_c$ of the HIP wire and tape under several magnetic fields at 4.2 K are summarized in figure 6 (a) and (b), respectively. The field directions applied to the samples are also described in figure 6(a) and (b). Transport $J_c$ in the HIP wire and tape from figure 2 are also plotted for comparison. In the HIP wire, the obtained magnetic $J_c$ well reproduces the transport $J_c$ [16,24]. The value of magnetic $J_c$ reaches approximately 200 kA/cm$^2$ which is twice the practical level of 100 kA/cm$^2$, at self-field.

In the magnetization measurements of the HIP tape, we observed an intriguing anisotropy of magnetic $J_c$. The magnetic $J_c$ has reached almost 430 kA/cm$^2$ at self-field when the magnetic field is parallel to the tape surface. This is the highest value among reported self-field $J_c$ in IBS wires and tapes. On the other hand, when the magnetic field is perpendicular to the tape surface, magnetic $J_c$ is substantially smaller compared with that for the field parallel to the tape surface. The magnetic $J_c$ in this configuration is almost 217 kA/cm$^2$ at self-field, which is smaller than the transport $J_c$ and approximately a half of the magnetic $J_c$ for field parallel to the surface. A similar anisotropy of magnetic $J_c$ have also been reported in pressed tapes of (Ba,K)Fe$_2$As$_2$ and (Sr,Na)Fe$_2$As$_2$ [42, 43]. This intriguing behavior can be explained qualitatively by the presence of weak-link area or microcracks in the core of the tape [43]. If there is a weak-link area along the tape length, the transport $J_c$ is not suppressed significantly. Existence of weak-link areas in the HIP tape has been confirmed in MO images of figure 5(d) and (e). In the presence of such a weak-link area, when the magnetic field is applied perpendicular to the tape surface, the flow of the shielding current is interrupted by the weak-link area as shown in Fig. 6(c). By contrast, when the magnetic field is parallel to the tape surface, the flow of the shielding current is little affected by the weak-link area as shown in Fig. 6 (d). This current direction is the same as transport current as shown in Fig. 6(c). In this case, reduction of magnetic $J_c$ should be minimal. This scenario explains the anisotropy of $J_c$ qualitatively. Actually, MO images clarified the presence of weak-link area in the HIP tape core as shown in figure 3 and 5. This weak-link area may interrupt the current flow when field is perpendicular to the surface as described in figure 6 (d). In this case, the magnetic $J_c$ is underestimated.

Analyses of MO images and magnetization measurements suggest the presence of inhomogeneous $J_c$ distribution in the core of the HIP tape. $J_c$ is significantly suppressed at the thinner part near the center of the gourd-shaped core. It is interesting that the $J_c$ in the HIP tape is higher than that in the HIP wire where $J_c$ shows more homogeneous distribution as shown in the MO image in figure 3(c). Clarifying the possible origin of such a difference should give beneficial information not only for tapes but also for wires. One possible reason for the smaller $J_c$ is too small thickness of the core. Gao *et al*. suggested that the value of $J_c$ in pressed tape is proportional to the Vickers hardness, which is related to density of the core [32]. They show that as the thickness of the core becomes thinner, $J_c$ becomes higher. However, they also show that too thin tape does not sustain the highest $J_c$. Such a behavior may be caused by the damage in the core. Another possibility is the nonuniform pressure in the core due to its gourd-like shape. Since our wires and tapes are drawn into a square shape during the PIT process, the shape of the core is not ideal. In this case, nonuniform pressure can be exerted when it is formed into a tape form. Furthermore, the effect of grain texture should be considered. Gao *et al*. also suggested that grain texture affects the enhancement of $J_c$. It is possible that distribution of $J_c$ value in the HIP tape is caused by inhomogeneous degree of grain texture caused by gourd-like shape. Anyway, our HIP tape shows relatively higher $J_c$ although the deformation process or pressing conditions have not been optimized yet. It suggests that there is a room for further improvement of $J_c$ properties by optimizing the PIT and successive deformation processes. Furthermore, the same argument can also be applied to the HIP wire. Higher density and more uniform pressure should be realized by optimizing the wire fabrication process, although texturing cannot be applied in the case of round wire. It is concluded that HIP method is another candidate for the fabrication of high $J_c$ tape rather than the pressed tapes.

**Conclusion**

In summary, we have prepared polycrystalline $(Ba,K)Fe_2As_2$ powders with improved quality and fabricated the HIP wires and tapes. They were characterized by magnetization and transport measurements, magneto-optical imaging, x-ray diffraction. We obtained the largest value of transport $J_c$ at 4.2 K, which reached 175 kA/cm$^2$ in self-field in a round wire. Furthermore, $J_c$ of the round wire at 100 kOe first exceeded 20 kA/cm$^2$. Improved synthesis of the polycrystalline powder may have played a key role in enhancing $J_c$. We also demonstrated an enhancement of $J_c$ in the tape by the HIP process. Magneto-optical imaging and magnetization measurements of the tape indicate that the $J_c$ in the tape core is inhomogeneous possibly due to non-ideal pressure during the tape formation. Optimization of the deformation process may lead to even larger $J_c$.

**Acknowledgement**

This work was partially supported by a Grant-in-Aid for Young Scientists (B) (No.16K17745), the Japan-China Bilateral Joint Research Project by the Japan Society for the Promotion of Science (JSPS), research grant of the Sumitomo foundation, and research grant (general research) of TEPCO memorial foundation. We thank Z. Gao, K. Togano and H. Kumakura for fruitful discussions for improvement of the quality of polycrystalline sample.

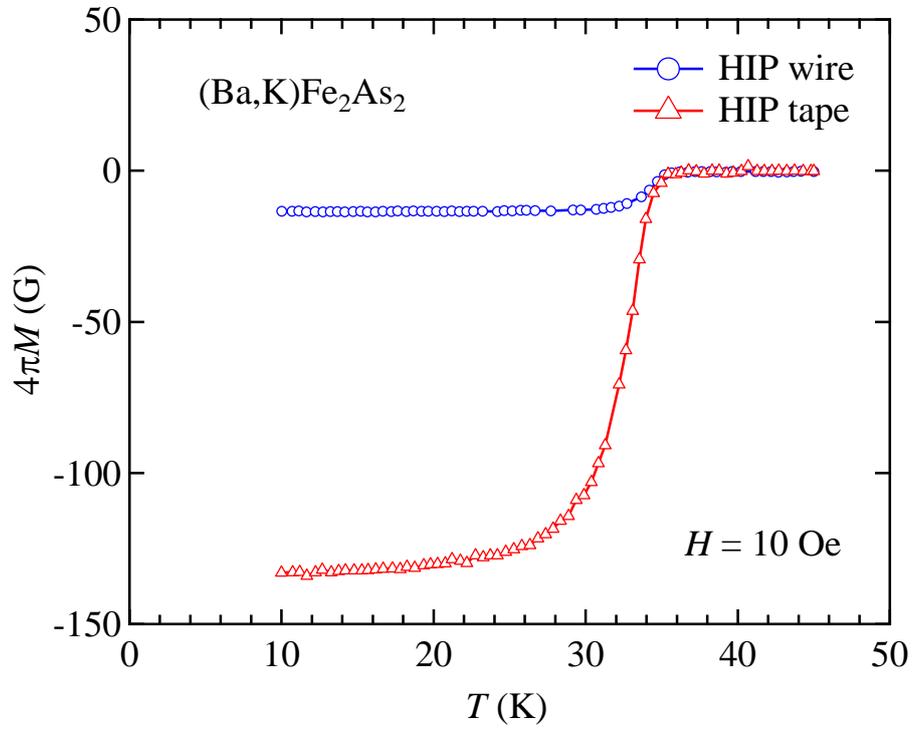

Figure 1. Temperature dependence of magnetization of the (Ba,K)Fe$_2$As$_2$ HIP wire and tape.

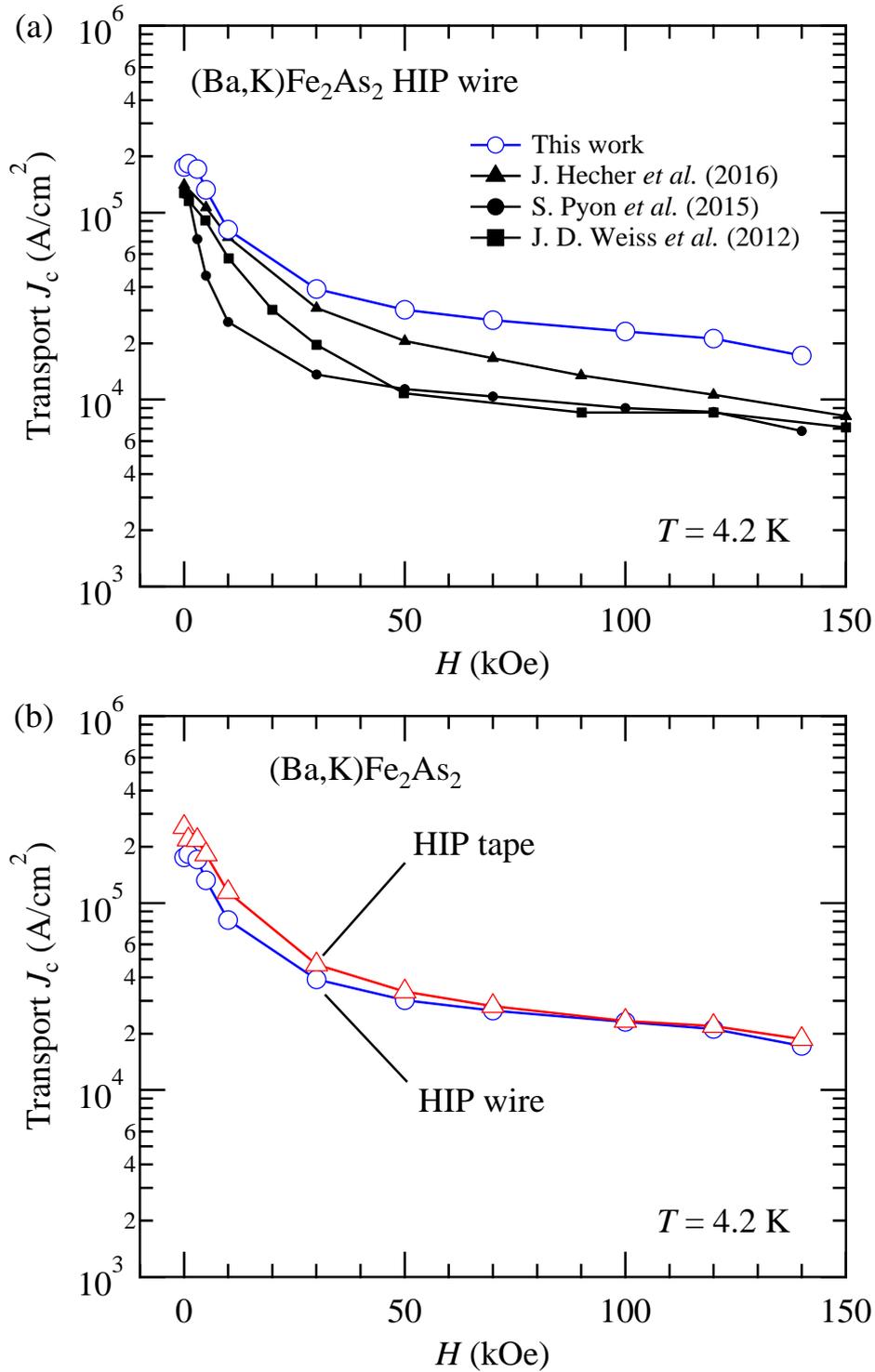

Figure 2. (a) Magnetic field dependence of transport $J_c$ in the $(Ba,K)Fe_2As_2$ HIP wire and tape at 4.2 K. Transport $J_c$ in the $(Ba,K)Fe_2As_2$ HIP wires from recent publications are also plotted [21,39,40]. (b) Comparison of magnetic field dependence of transport $J_c$ between the HIP wire and tape at 4.2 K.

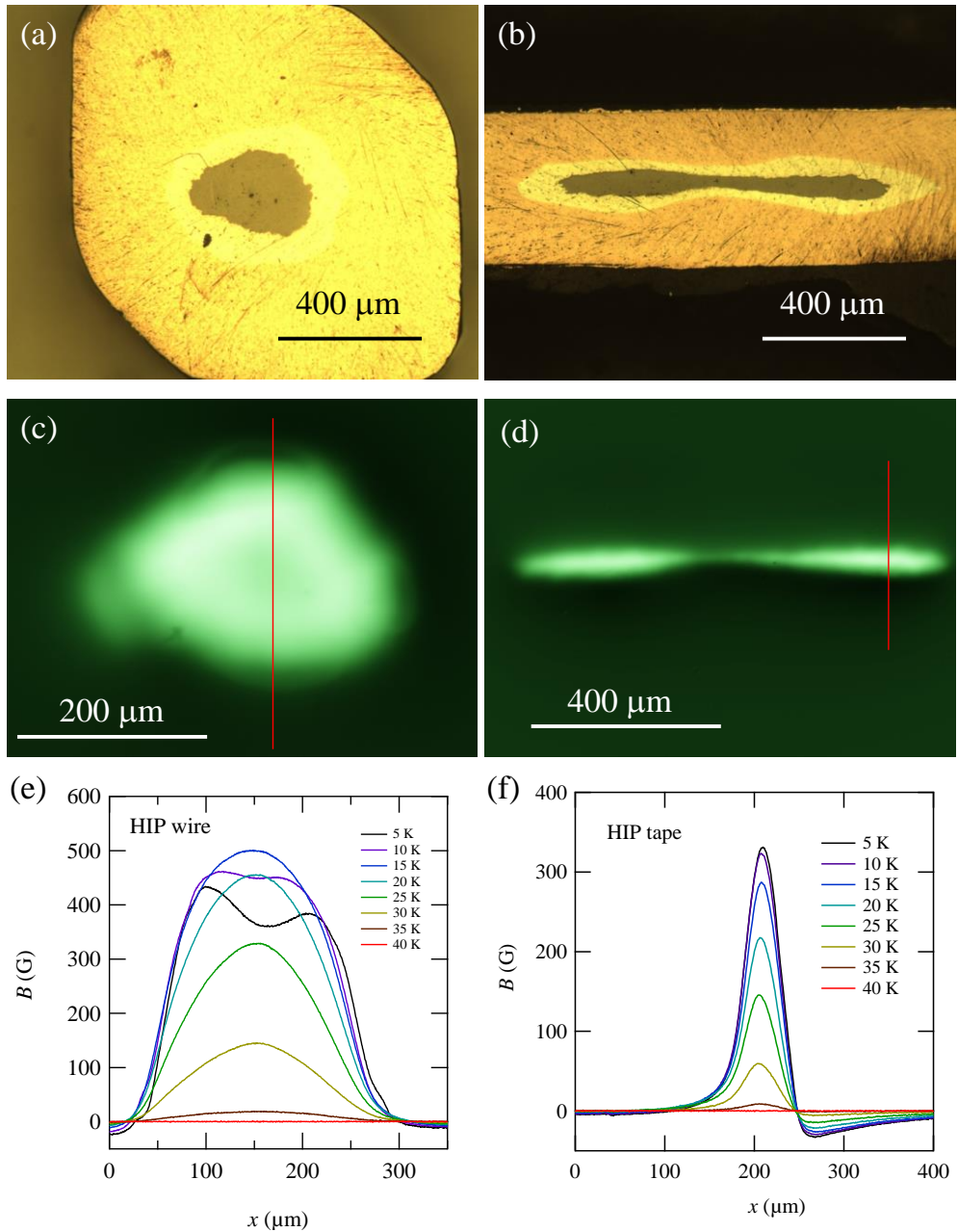

Figure 3. Optical micrographs of the transverse cross sections of the HIP (a) wire and (b) tape. Differential MO images of the core regions of (c) the HIP wire and (d) the HIP tape in the remanent state at 5 K after cycling the field up to 1.5 kOe for 0.2 s, which corresponds to (a) and (b), respectively. (e), (f) Local magnetic induction profiles at different temperatures taken along the red lines in (c) and (d), respectively.

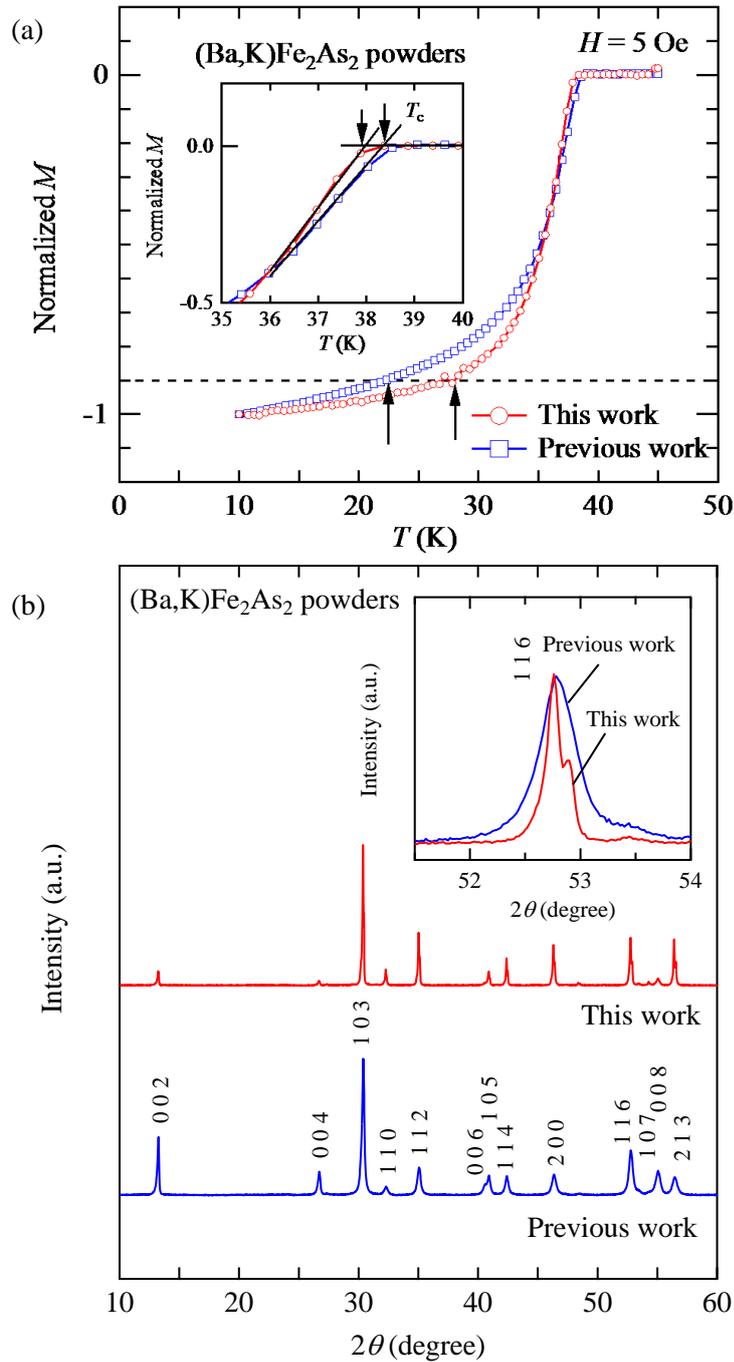

Figure 4. (a) Temperature dependence of normalized magnetization of $(Ba,K)Fe_2As_2$ powders (this work and previous work in ref. [39]). Dotted line and arrows indicate 90% values of magnetization at 10 K. The inset shows normalized magnetization near $T_c$. Arrows indicate the $T_c$ for the two powders. (b) Powder x-ray diffraction patterns of $(Ba,K)Fe_2As_2$ powders (this work and previous work in ref. [39]). The inset shows normalized (1 1 6) diffraction peaks of the two powders.

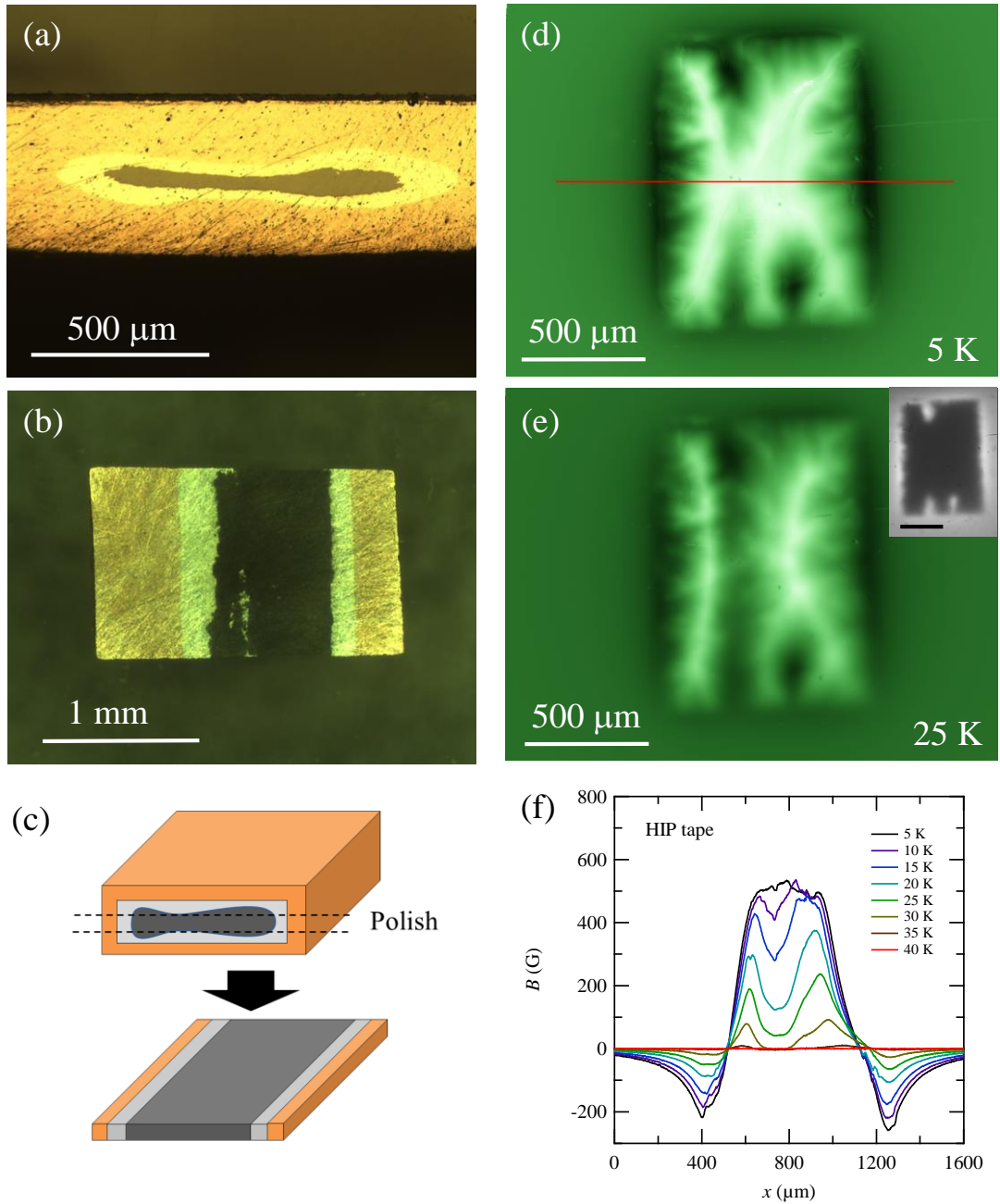

Figure 5. Optical micrographs of (a) the transverse cross section and (b) polished surface of the HIP tape. (c) Schematic drawing of the HIP tape before and after polishing. Differential MO images of the core region of the HIP tape in the remanent state at (d) 5 K and (e) 25 K, after cycling the field up to 1.5 kOe for 0.2 s. The inset of (e) is MO image of flux penetration in the core of HIP tape at 25 K and 50 Oe after zero-field cooling. The black bar in the figure corresponds to 500 μm. (f) Local magnetic induction profiles at different temperatures taken along the red line in (d).

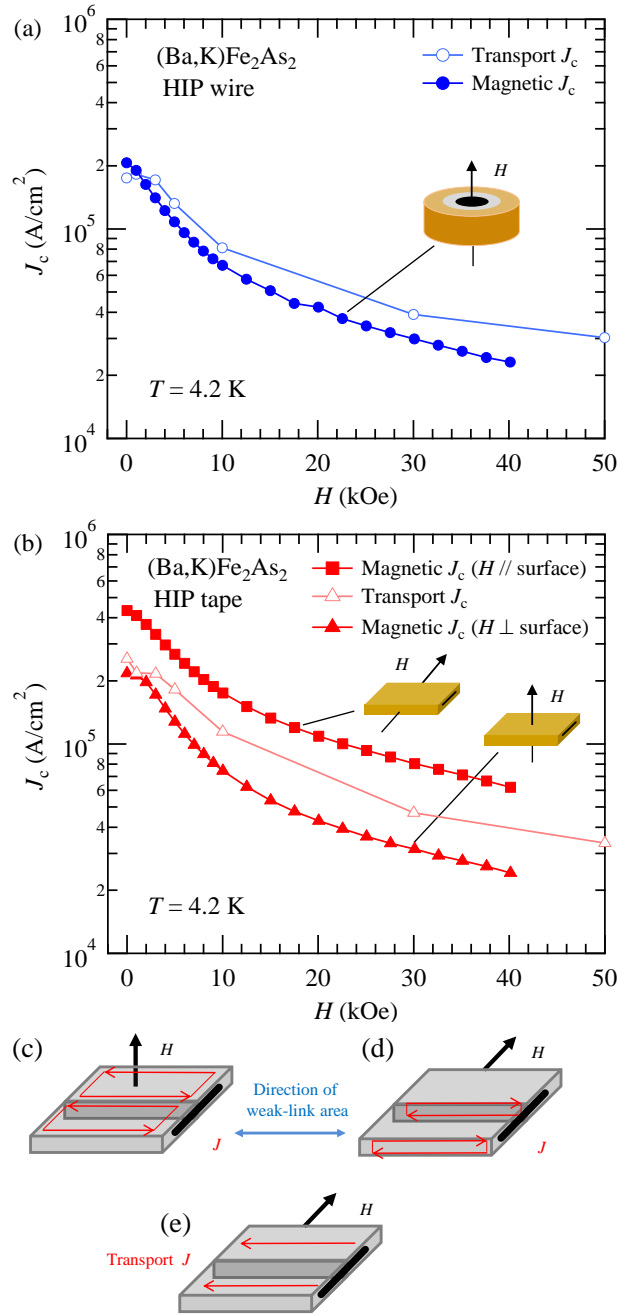

Figure 6. Magnetic field dependence of both transport and magnetic $J_c$ in the $(Ba,K)Fe_2As_2$ HIP (a) wire and (b) tape at 4.2 K. Magnetic field is applied parallel to the current flow direction for the HIP wire, and is applied parallel and perpendicular to the tape surface for the HIP tape as schematically drawn in the figure. (c)-(e) Schematic drawings of the relation between the magnetic field $H$ (black thick arrows) and the shielding or transport current $J$ (red arrows). Direction of the weak-link area is also shown by blue arrows.